# Characteristics of the Johari-Goldstein process in rigid asymmetric molecules

D. Fragiadakis and C.M. Roland

Naval Research Laboratory, Chemistry Division, Code 6120, Washington, DC 20375-5342

*(July 8, 2013)*

## ABSTRACT

Molecular dynamics simulations were carried out on a Lennard-Jones binary mixture of rigid (fixed bond length) diatomic molecules. The translational and rotational correlation functions, and the corresponding susceptibilities, exhibit two relaxation processes, the slow structural relaxation ($\alpha$ dynamics) and a higher frequency secondary relaxation. The latter is a Johari-Goldstein (JG) process, by its definition of involving all parts of the molecule. It shows several properties characteristic of the JG process – (i) merging with the $\alpha$ relaxation at high temperature; (ii) a change in temperature-dependence of the relaxation strength on vitrification; (iii) a separation in frequency from the $\alpha$ relaxation that correlates with the breadth of the $\alpha$ dispersion; and (iv) sensitivity to volume, pressure, and physical aging – that can be used to determine whether a secondary relaxation in a real material is an authentic JG process, rather than trivial motion involving intramolecular degrees of freedom. The latter has no connection to the glass transition, whereas the JG relaxation is closely related to structural relaxation, and thus can provide new insights into the phenomenon.

## INTRODUCTION

A full understanding, let alone a first principles model, of the dramatic slowing down of molecular motions in a vitrifying material remains a major unachieved goal of condensed matter physics. In addition to the primary mode of motion, the structural or α relaxation, glass-forming materials commonly show faster relaxation processes. Many of these are of intramolecular origin, and thus unrelated to structural relaxation and the glass transition. A particular type of secondary relaxation is the Johari-Goldstein (JG) process, which involves all atoms in the molecule and appears even in systems with completely rigid molecular structures [1]. The JG process seems to be universally present in glass forming materials, including molecular liquids, polymers, metallic glasses, and plastic crystals. There is a large amount of experimental evidence that the JG relaxation is closely related to or may even be the precursor of the α process.

Several decades after its discovery, the origin of the JG process remains unclear, and there are distinctly different hypotheses for the underlying mechanism; for example, do JG motions entail low amplitude reorientations of all molecules [2,3], or are they limited to those species in local regions of lower density [1]? Recent molecular dynamic (MD) simulations suggest the answer may depend on temperature and pressure [4]. Generally MD simulations hold great promise to investigate the JG relaxation by being able to address the simplest possible systems which capture the essential physics of the process. Many simulations of the glass transition have focused on mixtures of Lennard-Jones spheres. Although these exhibit local, thermally activated jump-like motions in the glass [5, 6], nothing resembling JG relaxations as seen experimentally appears in the dynamics of the simulated systems. A relaxation process with characteristics of the JG was observed by Bedrov and Smith in MD simulations of polybutadiene and a simple bead-chain polymer model [7,8]. Higuchi et al. [9,10] observed a secondary process in MD simulations of a flexible diatomic molecule, but only a weak indication was evident in the corresponding rigid molecule. In simulations of symmetric [11] and almost symmetric [12,13] dumbbell molecules with short bond lengths, 180° flips are prominent in the rotational dynamics; these enable the odd reorientational degrees of freedom to relax completely, even in the glassy state, while the even degrees of freedom remain frozen. These flips have some characteristics similar to secondary relaxations, but differ from experimental observations in glass-forming materials, where both first order (measured by dielectric spectroscopy) and second order (NMR or dynamic light scattering) rotational correlation functions only partially relax via the JG process, decaying to zero only over the longer α relaxation timescale. In a recent study, we found that this is due to the symmetry of the molecular structure; in an asymmetric diatomic molecule, a reorientational mechanism is observed that behaves much like the experimental JG process [4].

Here we simulate a family of asymmetric diatomic molecules, analogous to those of ref. [4], that exhibit a secondary β process. In experimental studies, a series of criteria have been proposed to distinguish the genuine JG process from secondary processes of intramolecular origin. In the case of a rigid molecule, the absence of intramolecular degrees of freedom guarantees that any secondary relaxation is a JG process, by its definition of involving the entire molecule. We test the β process of our simulated asymmetric diatomic molecules against these criteria, as well as various experimental correlations observed for the JG process to address the question: Is this β process the same phenomenon as the JG relaxation in real glass-forming liquids? In other words, how much of the physics of the JG relaxation, and its rich behavior observed experimentally, can be captured by our simple model system? This work will facilitate the continuing efforts to understand the nature of the JG process in glass forming materials.

## METHODS

Simulations were carried out using the HOOMD simulation package [14,15]. The systems studied are binary mixtures (4000:1000) of rigid, asymmetric diatomic molecules labeled AB and CD. Atoms belonging to different molecules interact through the Lennard-Jones potential

$$U_{ij}(r) = 4\varepsilon_{ij}\left[\left(\frac{\sigma_{ij}}{r}\right)^{12} - \left(\frac{\sigma_{ij}}{r}\right)^{6}\right] \qquad (1)$$

where $r$ is the distance between particles, and $i$ and $j$ refer to the particle types A, B, C and D. The energy and length parameters $\varepsilon_{ij}$ and $\sigma_{ij}$ are based on the Kob-Andersen (KA) liquid, a binary mixture that does not easily crystallize [16]. This was done as follows (noting that alternative choices of $\varepsilon_{ij}$ and $\sigma_{ij}$ give qualitatively similar results): The energy parameters $\varepsilon_{ij}$ are those of the KA liquid; i.e., $\varepsilon_{AA} = \varepsilon_{AB} = \varepsilon_{BB} = 1.0$, $\varepsilon_{CC} = \varepsilon_{CD} = \varepsilon_{DD} = 0.5$, and $\varepsilon_{AC} = \varepsilon_{AD} = \varepsilon_{BC} = \varepsilon_{BD} = 1.5$. To set $\sigma_{ij}$ we use the original KA parameters for the larger A and C particles, while the B and D particles are 50% smaller than A and C, respectively. Therefore, $\sigma_{AA} = 1$, $\sigma_{CC} = 0.88$, $\sigma_{BB} = 0.5$, and $\sigma_{DD} = 0.44$. For the particle interactions we take $\sigma_{ij} = S_{ij}(\sigma_{ii} + \sigma_{jj})$, where $S_{ij} = 0.5$ (additive interaction) when the particles belong to the same type of molecule ($I, j$ = AB, CD), and $S_{ij} = 0.4255$ when the particles belong to different types, the latter chosen to give the KA value for $\sigma_{AC} = 0.8$. All atoms have a mass of $m = 1$. The bond lengths A-B and C-D were fixed using rigid body dynamics [17]. All quantities are expressed in units of length $\sigma_{AA}$, temperature $\varepsilon_{AA}/k_B$, and time $(m\sigma_{AA}^2/\varepsilon_{AA})^{1/2}$.

A family of liquids with bond lengths d = 0.45, 0.5, 0.55, 0.6, and 0.7 were simulated. Except where noted, simulations were carried out in an NVT ensemble. Following ref. [11], densities were chosen to maintain a constant packing fraction; this results in a similar pressure range (approx. 0<P<10) for all molecules studied. The densities were $\rho$=1.3, 1.25, 1.21, 1.175, and 1.125, corresponding to the bond lengths above. Simulations at constant pressure give qualitatively identical results [4]. The time step was $\Delta t$ = 0.005. Data were collected at each temperature after an equilibration run several times longer than the structural relaxation time. At low temperatures, structural relaxation is extremely slow, and translational and orientational correlation functions do not decay to zero over the duration of the simulation runs; i.e., the system is out of equilibrium. For these conditions we increased the equilibration runs, until neither significant drift in volume nor aging of the translational and rotational correlation functions were observed; the residual rotational motion of the molecules at these temperatures takes place within a non-equilibrium, but essentially static structure.

The glass transition occurs in the simulations when the α relaxation time is much longer than the total (equilibration and production run) simulation time at a given temperature, which is on the order of $t_{max} \sim 10^6$. This is about 7 orders of magnitude slower than the vibrational relaxation times, so for an experimental glass-forming liquid would correspond to timescales in the range $10^{-5}$ s range, rather than the *ca.* 100 s for the usual experimental glass transition.

We follow the dynamics of the AB molecules (the behavior of the CD molecules is qualitatively the same). Rotational dynamics was studied via the first- and second-order rotational correlation functions calculated by

$$C_1(t) = \langle \cos\theta(t) \rangle$$
$$C_2(t) = \frac{1}{2}\langle 3\cos^2\theta(t) - 1 \rangle \qquad (2)$$

while translational dynamics were characterized via the center-of-mass self-intermediate scattering function $F_s(t)$ at a wavevector $q_{max}$ corresponding to the maximum in the static structure factor. The associated frequency-dependent susceptibilities were calculated via

$$\chi(\omega) = 1 + i\omega \int_0^\infty dt\, e^{i\omega t} \phi(t) \qquad (3)$$

where $\phi(t)$ is $C_1$, $C_2$ or $F_s$.

## RESULTS AND DISCUSSION

### Rotational relaxation

Figure 1 shows the first-order rotational correlation function for the AB molecules in the d=0.55 liquid for various temperatures. At short times, there is a small decrease in $C_1$ corresponding to oscillations within the local structure formed by neighboring particles, at a temperature independent $\tau_{vib} \cong 0.1$. At high temperatures $C_1$ then decays to zero via a single step. Below a temperature $T_{on}$, however, the relaxation occurs in two steps, a shorter time β and longer time α process. The latter appears as a long-time tail, which grows in strength with decreasing temperature at the expense of the β intensity. At T=0.4 and lower, the α relaxation time is much larger than the simulation run time; the system is in a non-equilibrium glassy state. Nevertheless, $C_1$ significantly relaxes to a non-zero value; the magnitude of this plateau increases with decreasing temperature.

The rotational dynamics can be more directly compared to experimental dielectric relaxation data by converting to the frequency-dependent susceptibility (Figure 1(b)). In this form the data show more clearly the change in response with decreasing temperature. Vibrational motion occurs at a temperature-independent frequency, followed by a broad, symmetric β peak, and a narrow, asymmetric α process that increases in magnitude on cooling.

Figure 2 shows the rotational correlation function and susceptibility for a system with a larger bond length, d=0.70. Again below an onset temperature, the spectrum is bimodal, although the separation of the two processes is smaller, and the onset temperature lower, than for the shorter molecule in Fig. 1. This bimodal character of the peaks is more readily apparent in the susceptibility spectra than in the time-correlation functions.

Higher-order rotational correlation functions, as well as translational relaxation, behave in a qualitatively very similar way, although the relative intensities of the α, β, and vibrational relaxations vary, and the relaxation times are slightly different. This can be seen in Figure 3, which shows the first- and second-order rotational correlation functions, along with the center-of-mass self-intermediate scattering function for a typical liquid state point.

**Deconvolution of α and β processes**

Determining the relaxation times, intensities, and shapes for the α and β processes requires deconvolution of the relaxation function ϕ(t) into the component functions $ϕ_α$ and $ϕ_β$. Two methods are commonly used to accomplish this: If the processes are independent and uncoupled [18,19], the relaxation functions are additive, so that the total relaxation function (excluding the vibrational contribution) can be described by

$$\phi(t) = \left(\Delta\phi_\alpha\right)\phi_\alpha(t) + \left(\Delta\phi_\beta\right)\phi_\beta(t) \tag{4}$$

where $Δϕ_α$ and $Δϕ_β$ are the α and β relaxation strengths, respectively. A second approach assumes the β process takes place in an environment that is rearranging on the timescale of the α process [20,21], with the two being "statistically independent"; this description yields the so-called Williams ansatz (WA) [22]

$$\phi(t) = \left(\Delta\phi_\alpha\right)\phi_\alpha(t) + \left(\Delta\phi_\beta\right)\phi_\alpha(t)\phi_\beta(t) \tag{5}$$

Both approaches are only approximate, and neither can be correct when there is significant overlap of the two dispersions because they are not independent, but rather correspond to motions of the same molecular units at similar time scales.

For $\phi_\alpha$ we use a stretched exponential function [23] for the primary relaxation, $\phi_\alpha(t) = \exp\left[-(t/\tau_\alpha)^{\beta_K}\right]$. When using the WA, the β process can be fit by a Cole-Cole function in the frequency domain, or its transform in the time domain. However, to obtain an acceptable fit using eq.(4), an asymmetric β peak is needed; we use the empirical Havriliak-Negami function [23]

$$\chi_\beta(\omega) = \left[1 + (i\omega\tau_{HN})^a\right]^{-b} \tag{6}$$

The two fitting methods can yield different results if there is significant overlap between the α and β processes. Figure 4 compares relaxation times derived from fits by either method, for molecules with bond lengths $d$=0.55 and $d$=0.70. For the former, the methods yield similar relaxation times, with a slightly faster α process for the WA at high temperature. The two processes behave according to the "splitting scenario", where a separate onset of the α process emerges at a temperature $T_{on}$ at which $\tau_\alpha$ and $\tau_\beta$ are significantly different; the high-temperature relaxation at T>$T_{on}$ appears as the continuation of the β process. For longer bond lengths, the WA gives again a separate onset for the α process, but with a very small separation of the two processes at the onset temperature. Fitting using the additive assumption gives a slightly different picture: the high-temperature relaxation appears as a continuation of the low-temperature α process, conforming to the so-called "merging scenario". This suggests that as the separation between the α and β relaxation times becomes very small (around a decade or less), the results depend on the (somewhat arbitrary) choice of fitting method .

### Dependence of relaxation behavior on bond length

Figures 5 and 6 show the variation with temperature of the relaxation times and strengths for the two processes, for bond lengths between 0.45 and 0.6. The β process shows Arrhenius behavior in the glassy state, while above $T_g$ some curvature in log τ vs. 1/T plots is evident. The β relaxation strength increases with increasing temperature, while that of the α decreases, going to zero at $T_{on}$. These are the same trends observed experimentally in the dielectric strength and relaxation times of supercooled liquids that conform to the "splitting scenario" for the α-β crossover region. With increasing bond length the β process slows down and its activation energy increases, as expected for a non-cooperative process (the potential barrier for local rotation of a single molecule should be higher for larger bond lengths). The behavior of the α process is more complex. With increasing bond length, the α dynamics becomes faster (lower $T_g$), until approximately d=0.65-0.7, whereupon the trend reverses, $T_g$ increasing with increasing d (not shown). The same behavior has been found in symmetric dumbbell molecules [11,24]. There it has been related to molecular packing: For non-spherical particles of various shapes such as ellipsoids [25] and spherocylinders [26], the maximum attainable packing fraction is a non-monotonic function of the aspect ratio; packing increases then decreases with increasing molecular elongation. The difference between the actual and maximum packing fraction, which determines the volume available for molecular reconfigurations and thus affects $T_g$, will increase and then decrease with increasing bond length. This underlies the observed dependence of $T_g$ on d.

## Testing for characteristic properties of the JG process

The purpose of this work is to determine if properties of the β process observed herein in MD simulations conform to various criteria proposed as characteristics of JG relaxations seen in experiment work [27,28].

**(a) Merging with the α process.**

An important characteristic of the JG process is that at high temperatures it merges with the α process [29, 30]. Such is the case here - relaxation functions for both rotational ($C_1$, $C_2$) and translational motions ($F_s$) exhibit merging; at high temperatures there is only a single decay of the correlation function, or equivalently one susceptibility peak. Note this observation requires asymmetric molecules, since for symmetric or quasi-symmetric diatomic molecules, odd-order rotational correlators show only local dynamics (i.e., single-molecule 180 degree flips), which do not appear in the even-order correlators. In symmetric dumbbell molecules, translational motion is also insensitive to the 180 degree flips because such flips leave the center of mass in the same position.

**(b) Change in temperature dependence of relaxation time and relaxation strength across $T_g$.**

On heating through the glass transition, the JG process generally shows a change in activation energy, along with a stronger temperature dependence of the relaxation strength [31,32,33] (non-JG relaxations sometimes show similar behavior [34]). For the simulated systems, the T-dependence of the secondary relaxation changes only monotonically with temperature (Fig. 5), without the marked change in activation energy reported for some liquids [31,32,33]. The relaxation strength, however, does show discontinuous behavior as the glass transition temperature is traversed (Fig. 6).

An empirical correlation has been reported between the activation energy $E_{JG}$ of the JG process in the glassy state and the glass transition temperature: $E_{JG}$ = 24k$T_g$ [35]. It has been pointed out that examining a wider range of materials reveals a wide spread of $E_{JG}/T_g$ around the mean value of 24 [36]. This correlation does not hold among the systems simulated here: increasing the bond length systematically decreases $T_g$, but the β activation energy increases. This is reminiscent of the behavior of *n*-alkyl methacrylates, where $T_g$ decreases with increasing alkyl chain length, but the JG activation energy is little affected. In the case of the methacrylates this has been ascribed to internal plasticization with increasing alkyl chain length causing the drop in $T_g$ [37]. We speculate that a similar effect may be active in the simulated system, caused by the decrease in $T_g$ due to increased packing efficiency.

**(c) Correlation of $\tau_\beta$ with width of the α process.**

It is an empirical observation that for a given value of $\tau_\alpha$, the JG relaxation time correlates with the breadth of the α dispersion, the latter usually quantified by the Kohlarusch stretch exponent $\beta_K$. Alternatively, this correlation implies that the ratio $\tau_\alpha/\tau_\beta$ increases as $\beta_K$ decreases [38]. These relationships show unambiguously a connection between the α and JG processes.

Figure 7 shows that the stretching exponent for the α process in our simulations is strongly correlated with the relative magnitude of the α and β relaxation times, $\tau_\alpha/\tau_\beta$. The inset shows that the β relaxation time is a function of $\beta_K$ at fixed $\tau_\alpha$ for the four systems studied. These correlations mirror experimental observations.

**(d) Effect of pressure.**

The JG relaxation time is sensitive to pressure, in contrast to the negligible pressure dependence of intramolecular secondary relaxations. In several materials it has been found that the α and β relaxations shift with pressure and temperature such that $\tau_{JG}$ is invariant at constant $\tau_\alpha$. Figure 8 shows the pressure dependence of the α and β relaxation times as a function of pressure at constant T=1.0, for two of the systems. The β process shows a strong pressure dependence, with activation volumes ($V^* \equiv RT \dfrac{d\ln\tau}{dP}$) of the same order of magnitude as those of the α relaxation: $V_\alpha^*$=0.067 and $V_\beta^*$=0.036 for the shorter molecule (d=0.5), and $V_\alpha^*$=0.079 and $V_\beta^*$=0.055 for d=0.6. Often experimental $V_\alpha^*$ are similar to the molecular volume $V_m$. For our simulated systems $V_\alpha^*$ is several times smaller than $V_m$; however, the experiments at high pressure are carried out to longer times and thus lower temperatures than herein, and $V_\alpha^*$ increases with decreasing temperature. For state points with the same $\tau_\alpha$ but densities differing by up to 20%, $\tau_\beta$ is only approximately constant, with a small but systematic speeding up of the β process with increasing P and T (not shown). At constant $\tau_\alpha$, the relaxation strength of the β process also systematically decreases with increasing (P, T).

**(e) Effect of physical aging.**

The intensity and relaxation time of the JG process in the glassy state are affected by physical aging, while other secondary processes are relatively insensitive to aging [39,40,41,42,43,44]. The relaxation strength typically decreases with aging time, and the relaxation time slightly decreases. The latter effect is counterintuitive, since physical aging is accompanied by an increase in density which, under equilibrium conditions, would increase $\tau_{JG}$.

Figure 9 shows the evolution of the relaxation time and strength of the β process during physical aging of the species with d=0.5. This simulation was carried out at a constant pressure P=1, with the liquid equilibrated at T=0.5, followed by an instantaneous temperature jump to the aging temperature (T=0.35, 0.4, or 0.45). A long NPT run was then carried out, and data collected at various times sufficiently long to observe the β process (t>$\tau_\beta$), but short enough that the change of the dynamics during each collection period was minimal. The potential energy (also shown in Figure 9) shows a marked decrease with aging time at fixed P and T, a clear signature of physical aging (the volume follows similar kinetics). The β relaxation strength decreases and the relaxation time slightly decreases as a function of aging time, similar to experimental observations.

The changes in the secondary relaxation caused by aging have been described in the framework of an asymmetric double well model [45,39]. This model is based on two quantities, the energy barrier U between the two wells and Δ, the energy difference between the two wells. Qualitatively, Δ is predicted to increase and U to decrease with aging time, and if the sum 2U+Δ remains constant during aging, the relaxation strength and relaxation time during aging should be related by a power law, $\Delta\phi_\beta \propto \tau_\beta^{0.5}$. In an experimental study of aging of polyvinylethylene, a power law was found with a smaller exponent of 0.34, reflecting a decrease of 2U+Δ with aging [39]. The present aging data are compatible with a power law exponent of 0.45±0.8 (Figure 10).

# SUMMARY

We carried out MD simulations on rigid diatomic molecules lacking internal degrees of freedom, whereby their secondary relaxations, intermediate between the temperature-insensitive vibrations and structural relaxation, are by definition the Johari-Goldstein type. Experiments on real liquids suggest that a number of properties serve as signatures of the Johari-Goldstein process; secondary relaxations exhibiting these properties are presumed to involve motion of all atoms in the molecule. These properties and the conformance of our simulated diatomics are:

1. At high temperatures only a single manifestation of either the translational or rotational dynamics is observed; thus, merging with the $\alpha$ relaxation is a characteristic of the JG process.

2. If the JG process serves as the precursor to structural relaxation, it is expected to "sense" $T_g$, and the T-dependence of the strength of JG relaxation does change as $T_g$ is traversed. However, no clear change in the activation energy is observed at $T_g$.

3. The separation in frequency of the JG and $\alpha$ relaxation peaks is determined by the breadth of the $\alpha$ dispersion. The implication is that the intermolecular cooperativity that broadens the $\alpha$ dispersion (dynamic heterogeneity more broadly distributing the $\alpha$ relaxation times) slows down the $\alpha$ process, moving it further away from the JG relaxation.

4. The JG relaxation is intermolecularly correlated and therefore sensitive to volume; this causes $\tau_{JG}$ to change with both pressure and physical aging.

Solving the glass transition problem is complicated by the many interrelated properties associated with vitrification of a liquid. The JG relaxation affords an opportunity to circumvent some of these complications, by serving as the precursor of structural relaxation while being less affected by intermolecular cooperativity. Progress requires correct identification of the JG relaxation among the myriad secondary relaxations exhibited by glass-forming materials, especially associated liquids and polymers. The work herein helps to clarify those properties that are inherent to the JG process.

## ACKNOWLDEGMENT

This work was supported by the Office of Naval Research.

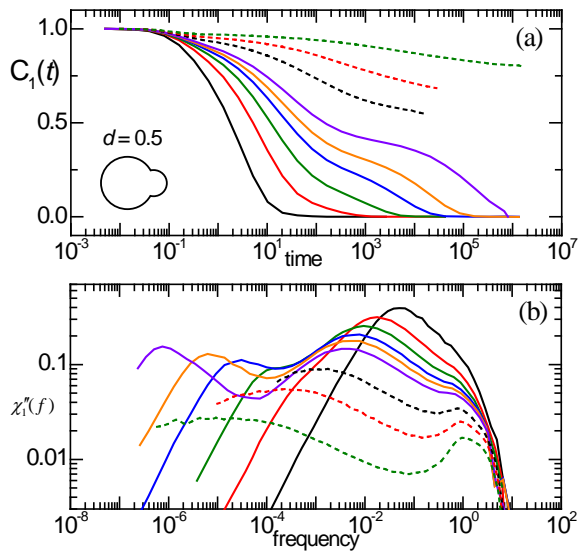

**Figure 1.** (a) First-order rotational correlation function and (b) imaginary part of the associated susceptibility, for the AB molecules in the system with *d*=0.5. Temperatures (short to long times, high to low frequencies) are T=1, 0.7, 0.6, 0.55, 0.52, 0.49 in the liquid state (solid lines) and T=0.4, 0.33, 0.25 in the glass (dashed lines).

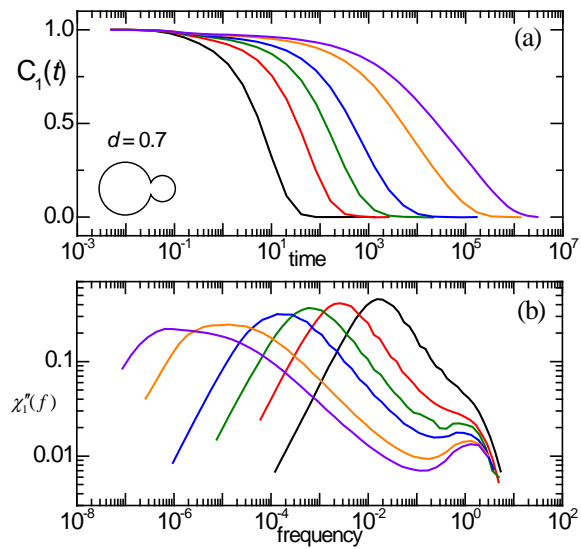

**Figure 2.** (a) First-order rotational correlation function and (b) imaginary part of the associated susceptibility, for the AB molecules in the system with d=0.7. Temperatures (short to long times, high to low frequencies) are T=1, 0.6, 0.5, 0.45, 0.40, 0.38.

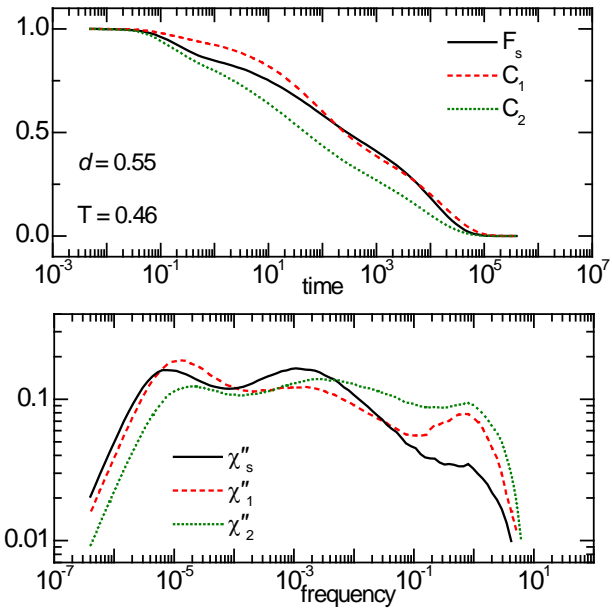

**Figure 3.** (a) Translational and rotational correlation functions for the AB molecules in the system with bond length $d$=0.55. Self-intermediate scattering function for the center of mass (solid line), first order (dashes) and second order (dots) rotational correlation function. (b) Imaginary part of the associated susceptibilities

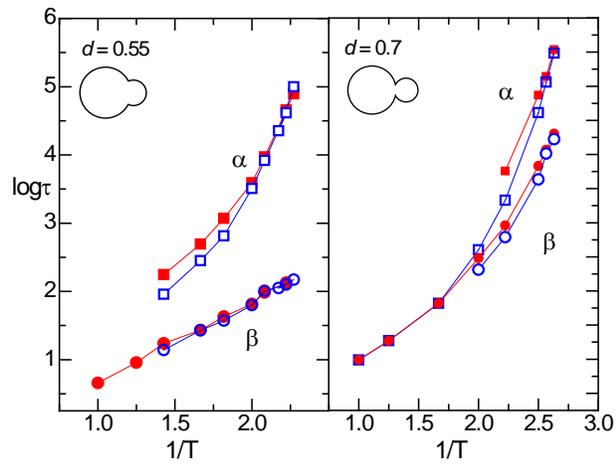

**Figure 4.** α- and β relaxation times for r=0.55 (short) and r=0.7 (long) molecules, obtained by fitting using the Williams ansatz (eq. (5), filled symbols) and additive combination (eq. (4), open symbols).

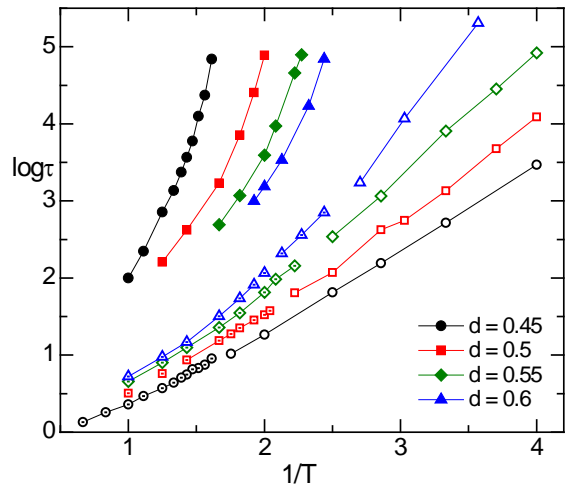

**Figure 5.** Temperature dependence of the rotational relaxation times of the α process (solid symbols) and β process (hollow symbols: glass; dotted symbols: liquid) for the systems with bond lengths d=0.45, 0.5, 0.55 and 0.6.

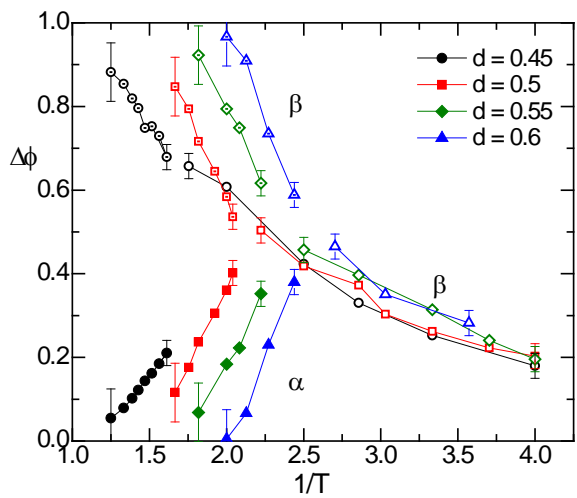

**Figure 6.** Temperature dependence of the intensities of the α (solid symbols) and β (hollow symbols: glass; dotted symbols: liquid) processes for the systems with bond lengths d=0.45, 0.5, 0.55 and 0.6.

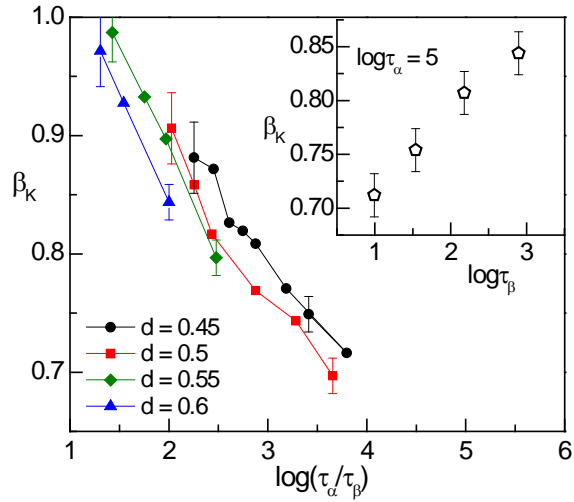

**Figure 7.** α-process stretch exponent as a function of the separation of α and β time constants for systems with bond lengths d=0.45, 0.5, 0.55, and 0.6. Inset: stretch exponent as a function of β relaxation time for the four systems at a constant α relaxation time of $10^5$.

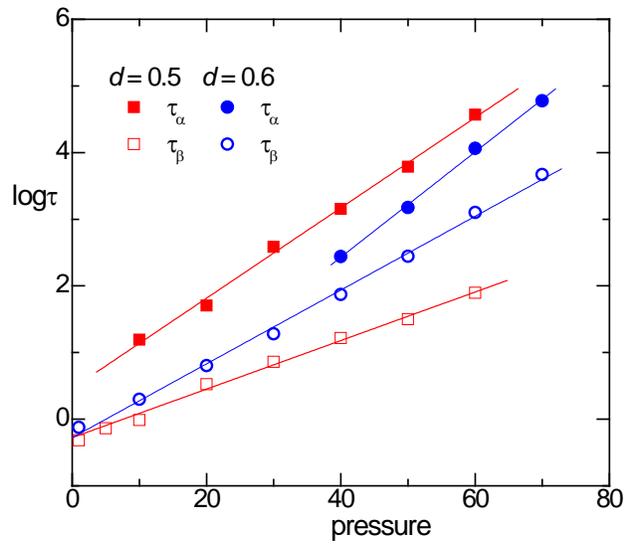

**Figure 8.** Relaxation times for α and β processes as a function of pressure for T=1.0, for the systems with bond lengths d=0.5 and d=0.6.

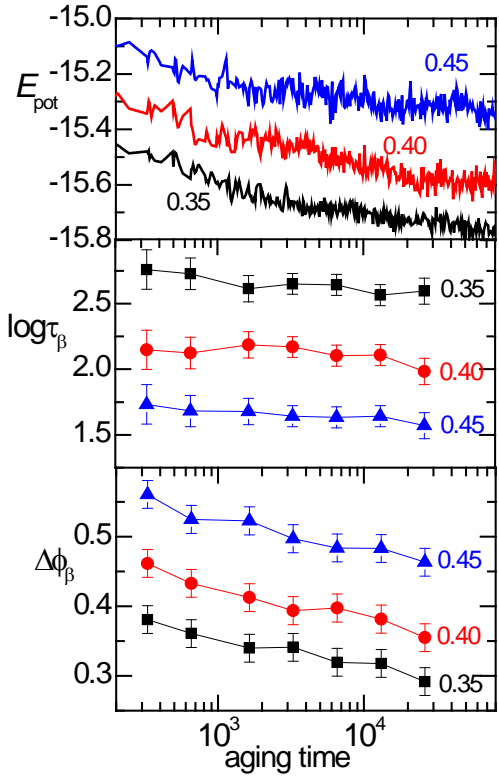

**Figure 9.** Physical aging in the glassy state, for bond length d=0.5: Potential energy per molecule (top), β relaxation strength (middle) and β relaxation time (bottom) as a function of aging time following a temperature jump from T=0.5 to temperatures T=0.45, 0.40, 0.35, under constant pressure P=1.

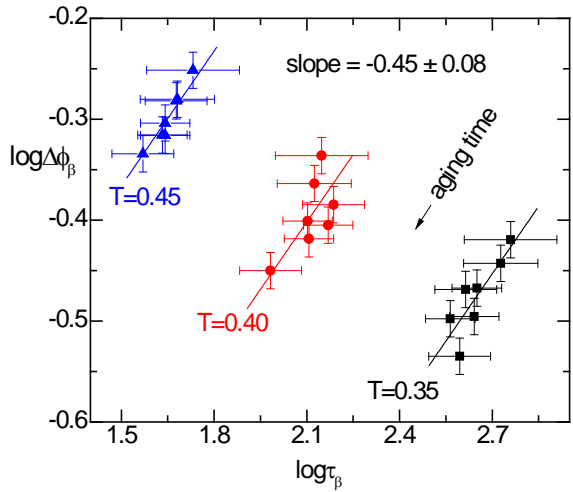

**Figure 10.** Relaxation strength vs. relaxation time during aging for bond length d=0.5. Lines are power law fits with a common exponent.